\documentclass{ws-procs9x6}

\begin{document}
\title{ Longitudinal polarization of $\Lambda$  and $\bar \Lambda$  
hyperons in deep-inelastic scattering at COMPASS. }
\author{ V.Yu.~Alexakhin \\
on behalf of the COMPASS Collaboration.}
\address{
Joint Institute for Nuclear Research \\ 
Dubna 141980, Russia\\
E-mail: Vadim.Alexakhine@cern.ch }
\maketitle
\abstracts{
The study of longitudinal polarization of $\Lambda$ ($\bar \Lambda$)
hyperons in deep-inelastic scattering is important because it
can provide an information
on the fundamental properties of the nucleon, such as
polarization of the strange quarks in the nucleon
and to determine the  mechanism of spin transfer from
polarized quark to a polarized baryon.
The production of  $\Lambda$ and $\bar \Lambda$  by
polarized $\mu^+$ of 160 GeV/c on a polarized $^{6}$LiD target has
been studied using the COMPASS spectrometer.
An important feature of the COMPASS experimental data sample is a large 
number of $\bar \Lambda$ hyperons, which is comparable with number of 
$\Lambda$. First preliminary results
on the longitudinal polarization of $\Lambda$ and $\bar \Lambda$
hyperons produced in the deep-inelastic scattering will be 
presented for 2002 data set.
}
Measurements of $\Lambda$ polarization in target fragmentation region
provide important information on the fundamental properties of the nucleon
such as the role of  the $\bar{s}s$ pairs in the proton wave function 
\cite{Ellis}.
The polarized nucleon intrinsic strangeness model \cite{Model,Model-2} 
predicts negative longitudinal polarization of $\Lambda$ hyperons
produced in target fragmentation region.
The measurement of the $\Lambda$ polarization in the current fragmentation
region allows to investigate another phenomenon, namely the spin
transfer from polarized quark to a polarized baryon. Recent
theoretical models of $\Lambda$ polarization in DIS
can be found in Refs.~\refcite{BJ}--\refcite{Soffer_group}. Another interesting topic
is study of possible quark-antiquark asymmetries either in the
quark to $\Lambda$ fragmentation functions
and/or in the quark and antiquark distributions of the target nucleon.
Calculations from model [\refcite{Soffer_group}] show that
spin transfer to $\Lambda$ and $\bar \Lambda$ should be the same if
standard quark distributions $s(x)=\bar s(x)$ are used. However there 
are difficulties in
the interpretation of the results due to large contribution from the
diquark fragmentation \cite{Model-2} and significant fraction 
of $\Lambda$ hyperons produced via decays of heavier hyperons.

The experimental situation of $\Lambda$ and $\bar \Lambda$ production
in DIS is summarized in Table~1. 
One can see that in the target fragmentation region
the $\Lambda$ polarization is negative, the spin transfer for current
fragmentation region seems to be small. 
\begin{table}[t]
\tbl{ Summary of experimental measurements of $\Lambda$ hyperon
  polarization in DIS. Sign of 
polarization is given with respect to virtual photon momentum.}
{\scriptsize
\begin{tabular}{lcccccc}
\hline
Reaction & $<E_{b}>$ &                &             &                 & & \\
Exp.                & (GeV)    & $x_F$ & $N_\Lambda$ & $P_\Lambda$  & $N_{\bar\Lambda}$ & $P_{\bar\Lambda}$   \\
\hline
$\bar {\nu_{\mu}}Ne$& 40       & $x_F<0$   & 403         & $-0.63\pm0.13$ &       &        \\
WA59\cite{wa59}      &          &$x_F>0$   & 66          & $-0.11\pm0.45$ &       &        \\
\hline
$\mu N$               &470   & \scriptsize{$0<x_F<0.3$} & 750 & $ 0.42\pm0.17$& 650   & $-0.09\pm0.20$  \\
E665 \cite{e665}     &          &$x_F>0.3$&                        & $0.09\pm0.19$&                 & $-0.31\pm0.22$  \\
\hline
$\nu_{\mu} N$         &43.8      & $x_F<0$  & 5608        & $ -0.21\pm0.04$ & 248     & $0.23 \pm 0.20$  \\
NOMAD \cite{NOMAD_l} &          & $x_F>0$  & 2479        & $ -0.09\pm0.06$ & 401     & $-0.23 \pm 0.15$  \\
\hline
$ eN$                 &27.5      & $x_F>0$ & 16900 &$\frac{P_{\Lambda}}{P_B D}=$ & 2500 & \\
HERMES \cite{hermes} &          &                &              & $0.06\pm0.09$   &         &         \\
\hline
\end{tabular}
}
\vspace*{-13pt}
\end{table}

COMPASS studies $\Lambda$ and $\bar \Lambda$  production by
polarized muons of 160 GeV/{\it c} on a polarized $^{6}$LiD target.
COMPASS spectrometer was constructed in the framework of CERN experiment 
NA58 [\refcite{Andrea}] .
The total amount of collected data during the run in 2002  is about 260 TB,
with typical event size of 35 kB. This analysis uses total 2002 
statistics obtained from longitudinally polarized target.
The data presented here are averaged on target 
polarization.

Selection criteria of $\Lambda$, $\bar\Lambda$ and $K^{0}_{s}$
($V^{0}$) are the following.
The primary vertex should be inside target cells whereas
the decay vertex of $V^0$ must be outside of the target.
The angle between vector of $V^0$ momentum and vector between
primary and $V^0$ vertices should be $\theta_{col}<0.01$ rad.
Cut on transverse momentum of the decay products with respect to the
direction of $V^{0}$ particle $p_t >23$ MeV/{\it c} was applied
to reject $e^{+}e^{-}$ pairs from the $\gamma$ conversion.
The standard DIS cut $Q^{2} >1~$ (GeV/{\it c})$^{2}$ and $0.2< y <
0.8$ have been used. After background subtraction  the experimental 
sample contains about 8000 $\Lambda$ and 5000 $\bar\Lambda$.
COMPASS is able to access mostly current fragmentation region with 
$<x_{F}>=0.2$, $<y>=0.45$, $<x_{Bj}>=0.02$ and $<Q^{2}>=2.62~(GeV/c)^2$.
The mean $\Lambda$ momentum is 12 GeV/{\it c}, while mean decay pion
momentum is 2 GeV/{\it c}. 
$\Lambda$ ($\bar\Lambda$) hyperon polarization can be measured via
asymmetry in the angular distribution of decay particles 
in $\Lambda \to p \pi^-$ ($\bar\Lambda \to \bar p
\pi^+$) decays. We determine X-axis along the direction of
the virtual photon in the $V^0$ rest frame. The angular distribution
in the $\Lambda$($\bar\Lambda$) rest frame is
\begin{equation}
\frac{dN}{d \cos{\theta_X}}=\frac{N_{tot}}{2}(1+\alpha P
\cos{\theta_X})
\end{equation}
where $N_{tot}$ is the total number of events, $\alpha=+(-)0.642\pm0.013$ is
$\Lambda$($\bar\Lambda$) decay parameter, $P$ is the
projection of the polarization vector on the corresponding axis,
$\theta_X$ is the angle between the direction of the
decay particle (proton for $\Lambda$, antiproton for $\bar\Lambda$)
and X-axis.
The acceptance correction was determined using
unpolarized Monte Carlo simulation. The DIS events are
produced by LEPTO 6.5.1 generator \cite{LEPTO} and the  apparatus is described by a full GEANT 3.21 \cite{GEANT} model. 

The analysis was performed slicing each angular distribution in 10 bins
and fitting the invariant mass distribution of  $V^0$ peak to obtain the number of
events in the bin. Corrected angular distributions were fit using
equation~(1). 

Results for longitudinal polarization are $P_{K^{0}_{s}}=0.007\pm0.017(stat.)$, $P_{\Lambda}=0.03\pm0.04(stat.)\pm0.04(syst.)$ and
$P_{\bar\Lambda}=-0.11\pm0.06(stat.)\pm0.05(syst.)$. All results were
averaged over accessible kinematical region.  

Spin transfer variable from beam to $\Lambda$ was calculated as 
$S=\frac{P_{\Lambda}}{P_B D}$, where $P_B$ is beam polarization and 
$D=(1-(1-y)^{2})/(1+(1-y)^{2})$ is virtual photon depolarization
  factor. Spin transfer for world and COMPASS
data are presented in Fig. 1 ( $\Lambda$ hyperons) and Fig. 2 
( for $\bar\Lambda$ hyperons). Only points with $x_{F}>0$ are shown.
One can see that there is reasonable agreement between COMPASS and
world data. There is indication that spin transfer to $\bar\Lambda$ is 
non-zero and might be different from the $\Lambda$ one.  
\begin{figure}[ht]
\centerline{\epsfxsize=2.5in\epsfbox{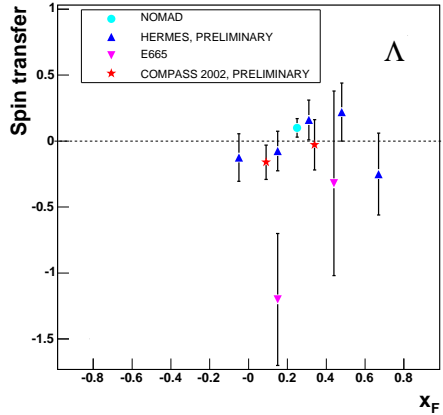}}
\caption{ Spin transfer for $\Lambda$ hyperons.}
\vspace*{-13pt}
\end{figure}
\begin{figure}[ht]
\centerline{\epsfxsize=2.5in\epsfbox{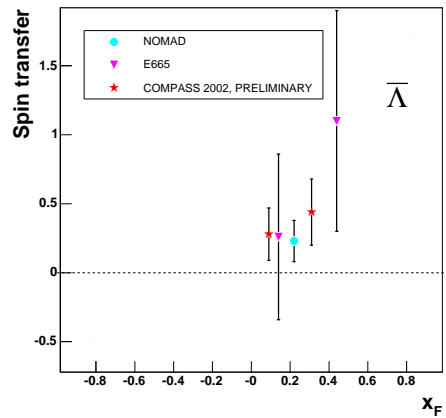}}
\caption{ Spin transfer for $\bar \Lambda$ hyperons.}
\end{figure}

COMPASS 2002 data show good potential for $\Lambda$ and $\bar\Lambda$ 
hyperons polarization measurement.
Data samples collected in 2003 and 2004 will significantly increase
the statistics.


\begin{thebibliography}{0}
\bibitem{Ellis} J.Ellis et al,
Phys.Lett., {\bf B353} (1995) 319; Nucl.Phys., {\bf A673} (2000) 256
\bibitem{Model}
J.~Ellis, D.~Kharzeev, A.M.~Kotzinian,  Z.Physik {\bf C69} (1996) 467
\bibitem{Model-2}
  J.~Ellis, A.M.~Kotzinian and D.V.~Naumov,  Eur. Phys. J., {\bf C25} (2002) 603
\bibitem{BJ} M.~Burkardt and R.~L.~Jaffe, Phys. Rev. Lett. {\bf 70} (1993)  2537
\bibitem{Anselmino_group}
  M.~Anselmino, M.~Boglione, U.~D'Alesio, E.~Leader and F.~Murgia,
  Phys.\ Lett.\ B {\bf 509} (2001) 246,
\bibitem{Kotzinian_group}
  A.~M.~Kotzinian, A.~Bravar and D.~von Harrach,
  Eur.\ Phys.\ J.\ {\bf C2} (1998) 329,
\bibitem{Boros}
  C.~Boros and L.~Zuo-Tang,
  Phys.\ Rev.\ {\bf D57} (1998) 4491.
\bibitem{Soffer_group}
  B.~Q.~Ma, I.~Schmidt, J.~Soffer and J.~J.~Yang,
  Phys.\ Lett.\ {\bf B488} (2000) 254
\bibitem{Andrea} A.~Bressan, contribution to these proceedings.
\bibitem{LEPTO} G.~Ingelman, A.~Edin and J.~Rathsman
Comp.\ Phys.\ Comm.\  {\bf 101}  (1997) 108
\bibitem{GEANT} GEANT, CERN Program Library Long Writeup W5013
\bibitem{wa59}
Willocq~S. {\it et al.} [WA59 Collaboration],
Z.Phys.  {\bfseries C53}, (1992) 207
\bibitem{e665}
  M.~R.~Adams {\it et al.}  [E665 Collaboration],
  Eur.\ Phys.\ J.\ {\bf C17} (2000) 263
\bibitem{NOMAD_l}
  P.~Astier {\it et al.}, [NOMAD Collaboration],
  Nucl. Phys. {\bf B588} (2000) 3;\\
  P.~Astier {\it et al.}  [NOMAD Collaboration],
  Nucl.\ Phys.\ {\bf B605}  (2001) 3
\bibitem{hermes}
  A.~Airapetian {\it et al.}  [HERMES Collaboration],
  Phys.\ Rev.\ {\bf B64} (2001) 112005
  S.~Belostotski, talk at this conference.
\end{thebibliography}
\end{document}